\documentclass[12pt]{article}
\usepackage{amsmath}
\usepackage{graphicx,psfrag,epsf}
\usepackage{enumerate}
\usepackage{natbib}
\usepackage{authblk}
\usepackage{url}
\newcommand{\blind}{0}
\begin{document}
\def\spacingset#1{\renewcommand{\baselinestretch}
{#1}\small\normalsize} \spacingset{1}
\if0\blind
{
\title{\bf A network flow approach to visualising the roles of covariates in random forests}
\author[,1,2,3,4]{Benjamin R. Fitzpatrick\thanks{The work has been supported by the Cooperative Research Centre for Spatial Information (CRCSI), whose activities are funded by the Australian Commonwealth's Cooperative Research Centres Programme. BRF gratefully acknowledges receipt of an Australian Post Graduate Award from the CRCSI.}}
\author[1,2,3,4]{Kerrie Mengersen}
\affil[1]{\small Mathematical Sciences School, Queensland University of Technology (QUT), Brisbane, QLD 4001, Australia}
\affil[2]{\small Cooperative Research Centre for Spatial Information (CRCSI), Carlton, VIC 3053, Australia}
\affil[3]{\small Institute for Future Environments, Queensland University of Technology (QUT), Brisbane, QLD 4001, Australia}
\affil[4]{\small ARC Centre of Excellence for Mathematical and Statistical Frontiers, Queensland University of Technology (QUT), Brisbane, QLD 4001, Australia}
  \maketitle
} \fi
\if1\blind
{
  \bigskip
  \bigskip
  \bigskip
  \begin{center}
    {\LARGE\bf Title}
\end{center}
  \medskip
} \fi

\bigskip
\begin{abstract}
We propose novel applications of parallel coordinates plots and Sankey diagrams to represent the hierarchies of interacting covariate effects in random forests.
Each visualisation summarises the frequencies of all of the paths through all of the trees in a random forest.
Visualisations of the roles of covariates in random forests include: ranked bar or dot charts depicting scalar metrics of the contributions of individual covariates to the predictive accuracy of the random forest; line graphs depicting various summaries of the effect of varying a particular covariate on the predictions from the random forest; heatmaps of metrics of the strengths of interactions between all pairs of covariates; and parallel coordinates plots for each response class depicting the distributions of the values of all covariates among the observations most representative of those predicted to belong that class.
Together these visualisations facilitate substantial insights into the roles of covariates in a random forest but do not communicate the frequencies of the hierarchies of covariates effects across the random forest or the orders in which covariates occur in these hierarchies.
Our visualisations address these gaps.
We demonstrate our visualisations using a random forest fitted to publicly available data and provide a software implementation in the form of an R package.
\end{abstract}

\vfill

\newpage
\spacingset{1.45} 
\section{Introduction}
\label{sec:intro}
Random forests \citep{Breiman2001} have achieved substantial popularity in the data mining and machine learning communities as a modeling method with good performance that is relatively straightforward to tune.
Random forests consist of many decision trees each fitted to a separate bootstrapped sample of the training data.
These trees are grown by dividing each current terminal node on the basis of one of a random sample of the covariates.
An independent random sample of the covariates is drawn for each node to be divided.
The division of the observations in a terminal node into two daughter nodes is chosen as that which most reduces the value of some loss function such as the sum of the squared errors associated with the predictions.
Each tree is grown in this manner until a stopping criterion is satisfied at which point no further divisions of the terminal nodes are attempted.
Stopping criteria include a minimum number of observations in any terminal node, a maximum number of nodes in a tree or an entropy threshold.
Trees are grown for typically hundreds to thousands of bootstrap samples of the data resulting in an ensemble of decision trees collectively termed a random forest.
The prediction of an observation with a random forest is calculated as the average (or majority vote) of the predictions of this observation with each tree in the random forest.
Thus while the process of fitting a random forest is relatively comprehensible, the resulting large ensemble of decision trees makes gaining a complete understanding of the roles of covariates in this model challenging.
\newline
\newline
Understanding the roles of covariates in a random forest is worthwhile even if producing an accurate and generalisable model is the only objective of the modelling.
This is because such understanding enables assessment of the model in terms of the current understanding of the system being modelled and a model that better reflects the nature of the system being modelled should generalise more effectively.
However, modelling objectives often include both insight into the system being modelled and accurate predictions.
Furthermore, models that are not interpretable in terms of the current understanding of the system being modelled may be distrusted regardless of their predictive performance.
Visualising a model can assist with such interpretation along with gaining and communicating such insights.
\newline
\newline
\citet{Wickham2015} explain the importance of both `visualisations of the data in the model space' and `visualisations of the model in the data space'.
The canonical example of visualisations of the data in the model space are residuals versus fitted values plots.
While important for model diagnostics, these plots do not facilitate insights into the roles of covariates in a model.
Visualisations of a model in the data space can facilitate insight into the roles of covariates through depiction of the responses of the model to features of the covariates. 
The high dimensions of the data to which random forest are typically applied has lead the various visualisations of random forests in the associated data spaces to include graphs of the predictions obtained across subspaces or projections of the data space.
Such plots typically depict the marginal effects of one or two covariates per plot and do not permit a holistic understanding of the roles of all covariates in the random forest.
Another key recommendation by \citet{Wickham2015} is to visualise collections of models rather than just exemplar models with high predictive performance.
In the context of visualising random forests this can be interpreted as a recommendation to visualise all trees constituent to a particular random forest.
The recommendation to visualise collections of models could also be interpreted as recommending the comparison of visualisations produced for multiple random forests each fitted to the same data using different values for the tuning parameters.
The tuning parameters for a random forest include: the size of the random sample of covariates from which a covariate is selected to define a node, those controlling the particulars of the bootstrap sampling and the parameter controlling the stopping criterion.
\newline
\newline
In this paper we propose novel visualisations of the roles of covariates in random forests.
We have designed these visualisations to be holistic representations of the structures of random forests that augment existing visualisations of the roles of covariates in random forests.
The visualisations we propose are novel applications of parallel coordinates plots and Sankey diagrams that represent all possible paths through the decision trees that constitute a random forest.
Our visualisations communicate the identities and orders of the sequences of covariates that define paths through the decision trees of a random forest.
Furthermore we have designed these visualisations to foreground the paths most frequently selected across all the decision trees that constitute a random forest.
In this way we seek to communicate all of the sequences of interacting covariates that are important to the predictive mechanism of a random forest in a single plot.
In Section \ref{sec:review.rf.vis} we review existing visualisations that facilitate insight into the roles of covariates in a random forest and introduce the visualisations we propose in this context.
The construction and interpretation of the visualisations we propose are best explained in terms of an example so we introduce a publicly available data set and describe how we fitted a random forest to these data in Section \ref{sec:example.rf}.
In Section \ref{sec:pcp} we explain how we apply parallel coordinates plots to represent the paths through the decision trees of our example random forest.
In Section \ref{sec:sankey} we explain the how we represent the same information with Sankey diagrams.
We open Section \ref{sec:disc} by discussing how our visualisations complement the existing collection of visualisations for investigating the roles of covariates in random forests.
We then outline the new insights our visualisations enable that these other visualisations do not.
We conclude by proposing elaborations upon our visualisations that could be attempted in future work.
A software implementation of our visualisation methods is provided via GitHub in the form of an R \citep{R2017} package.
Details regarding how to access this package are provided in the Supplementary Materials

\section{Visualising the Roles of Covariates}
\label{sec:review.rf.vis}
The simplest visualisations of the roles of covariates in a random forest are dot or bar charts that rank covariates by some metric of importance.
One such metric relates to the accumulation of the improvement in the splitting criterion across all nodes in the random forest at which the split was defined by the covariate in question \citep{Hastie2009.15.3.2}.
Another such metric is calculated from the change in the accuracy with which out of bag samples are predicted when the values of the covariate in question are permuted among them \citep{Hastie2009.15.3.2}.
Covariate importance may also be quantified on the basis of the variation in the predictions obtained from a sensitivity analysis conducted with respect to that covariate \citep{Cortez2013}.
The feature contribution method may be applied to random forests fitted for classification to quantify the importance of the covariates for predicting observations of each class in turn \citep{Palczewska2014}.
Summary statistics (e.g. medians) of the contribution of each covariate to prediction of the observations in each class may be visualised as grouped bar charts.
\newline
\newline
While scalar metrics of the importance of individual covariates are a useful starting point, they give little indication of the nature of the roles these covariates play in the random forest beyond the importance of that role.
Further insights into the nature of the roles of individual covariates in a random forest may be sought from a family of visualisations which summarise or project the predicted surface for two or three dimensional plotting.
Two examples of these sorts of plots are partial dependence (PD) plots \citep{Friedman2001} and variable effect characteristic (VEC) curves \citep{Cortez2013}.
The two dimensional versions of both types of plots depict a summary of the effect of a covariate of interest upon the predicted values as a single line.
Partial dependence (PD) plots may be applied to a variety of ensemble learning methods including random forests.
The most frequently encountered examples are two dimensional plots with the covariate of interest mapped to the horizontal axis and the vertical axis representing either the predicted value in the case of regression or a function of the predicted probability for a response class in the case of classification.
The curve in a PD plot is the result of averaging a collection of curves.
Each of these curves is a projection of the predicted surface onto a plane defined by the axes for predicted values and the covariate of interest.
Each of these projections is produced for the vector of covariate values (excluding the value of the covariate of interest) associated with one of the observations in the training set.
Three dimensional analogues are also possible with two covariate axes and one axis for the predictions.
\newline
\newline
Variable effect characteristic curves may be produced as part of a sensitivity analysis for a model \citep{Cortez2013}.
A sensitivity analysis involves examining the variation in predicted values obtained from a model supplied with a range of covariate values and thus is not specific to any particular model structure \citep{Cortez2013}.
The simplest form of sensitivity analysis examines the sensitivity of the model to a single covariate.
Vectors of covariate observations are constructed by combining each of a sequence of values of the covariate of interest with a set of values for the other covariates which are central to their distributions (e.g. the respective means or medians of these covariates over the training data).
These vectors are then used to predict the response with the model.
The resulting predictions are plotted against the associated values of the covariate of interest in what \citet{Cortez2013} term a variable effect characteristic (VEC) curve.
In the case of models fitted for classification tasks one such plot can be produced for the predicted probability of each response class.
The motivation of a VEC curve is similar to that of a PD plot.
A VEC curve connects a single sequence of predictions from the model where the only variation in the covariate vectors used to calculate these predictions is in the value of the covariate of interest.
In contrast, the curve in the PD plot is the average of many such curves produced at each of the set of covariate values associated with an observation in the training data.
While depicting overall summaries of the effect of the covariate of interest on the predicted values, any interaction between this covariate and another will not be visible from two dimensional PD plots and VEC curves.
All the variation among the parallel projections at different values of the other covariates is compressed into the single average curve displayed in the PD plot and in a VEC curve we see only a single such projection.
While the three dimensional analogues of these plots address this concern, they can only display evidence of two way interactions and random forests are ensembles of decision trees that describe interactions between numerous covariates.
\newline
\newline
Sensitivity analyses may be conducted to investigate the sensitivity of a model to any number of covariates simultaneously \citep{Cortez2013}.
First, a regular lattice is created which contains all combinations of sequences of values of each of the covariates under investigation.
Complete vectors of covariate values are then created by combining the covariate values associated with each point in lattice with a vector containing the central values of all the covariates not under investigation.
The model is then used to predict the response from each of these covariate vectors.
Two dimensional visualisations can then be constructed for each covariate of interest in turn.
In such visualisations the covariate of interest is mapped to the horizontal axis and the values predicted from the model mapped to the vertical axis.
The distributions of predicted values obtained from the sensitivity analysis at each value of the covariate of interest are represented as boxplots or intervals on these plots.
Three dimensional analogues of these plots can be constructed as heatmaps or contour plots with one covariate of interest on either axis and a summary statistic of the dispersion of predicted values from the sensitivity analysis mapped to the third dimension or colour in the plot.
\newline
\newline
Like PD plots and VEC curves, individual conditional expectation (ICE) plots \citep{Goldstein2015} display the predicted values from a model on the vertical axis and the value of a covariate of interest on the horizontal axis.
However, rather than displaying an average or interval of predicted values, an ICE plot displays separate curves for the projection of the predicted surface onto these axes at each of the covariate values available in the training data (or a subset thereof).
In this way ICE plots are somewhat akin to a data based sensitivity analysis where all covariates are varied.
Vertically centred ICE plots and a version that plots the partial derivative of the predicted surface with respect to the covariate of interest on the vertical axis are also available.
The curves (or sequences of points) in these plots may be coloured proportionally to the value of a second covariate of interest to search for evidence of two way interactions among the covariates.
\newline
\newline
Among the plots produced by the Random Forest Tool (RAFT) \citep{Breiman2004} is a heatmap depicting a metric of the strengths of interactions between all possible pairs of covariates.
These heatmaps use a metric of the strength of an interaction between covariates $x_i$ and $x_j$ defined on the basis of how much more or less likely a decision tree is to contain a binary partition defined using the covariate $x_i$ if above it in the tree is a binary partition defined using the covariate $x_j$.
While such heatmaps of sensitivity metrics for pairs of covariates allow for rapid inspection of many pairs of covariates for evidence of interactions such assessments rely on the quality of the scalar metric of pairwise sensitivity employed.
In such situations it would be advisable to produce multiple heatmaps each making use of a different sensitivity metric.
Insight into the nature of any interactions detected from these heatmaps would require pairwise plots such as coloured or three dimensional ICE, VEC or PD plots to be produced and inspected for each pair of interacting covariates identified from these heatmaps.
The combinatorially large number of potential interactions between even relatively moderate numbers of covariates renders assessment of the nature of each potential interaction by inspection of pairwise plots a time consuming option.
Furthermore, such an approach can only detect pairwise interactions.
Random forests are ensembles of many decision trees, each of which has the potential to model interactions between many covariates.
Consequently, a thorough understanding of the roles of covariates in a random forest will require examination of the potential for interactions of much high order than pairwise interactions.
Thus, visualisations of the entire model in the data space that highlight important interactions would seem to be promising tools for identifying important interactions.
\newline
\newline
The high dimensionality of the data to which random forests are often fitted make visual representation of the entire data space challenging.
Parallel coordinates plots of the entire data space and multiple dimensional scaling (MDS) plots summarising the entire data space have been used for this task.
Linked collections of distinct visualisations organised into interactive dashboards are a method multiple authors have adopted to summarise and convey the large volumes of information involved in visualising a random forest in the data space. 
The original example of an interactive dashboard for visualising random forests is the Java based random forest tool (RAFT) \citep{Breiman2004}.
The RAFT is only available for random forests fitted for classification.
\citet{Quach2012} developed another such interactive dashboard using R and the R package iPlots eXtreme \citep{Urbanek2011} for the computational performance improvements it permitted over the RAFT.
\citet{daSilva2017} have created a system of linked interactive graphics for diagnosis and exploration of an ensemble classifier which includes some comparisons to random forests fitted to the same data.
This dashboard was produced with R, ggplot2 \citep{Wickham2009}, plotly \citep{plotly2017} and shiny \cite{shiny2017} and includes visualisations of the models in the data space.
The visualisations of the RAFT, \citet{Quach2012} and \citet{daSilva2017} are useful aids for the discovery of clusters of observations among the training data and understanding covariate importance within these clusters.
\newline
\newline
Both \citet{Breiman2004} and \citet{Quach2012} use MDS plots to represent the observations in a subspace of the data space.
These MDS plots use principal components calculated from the proximity matrix of the random forest being visualised.
The $i$th row of the $j$th column of the proximity matrix is the proportion of trees in the random forest in which the pair of observations indexed by $i$ and $j$ were assigned to the same terminal node (and thus predicted to have the same response value).
In both dashboards observations selected from the MDS plot are then highlighted in the other plots available in these dashboards.
In this manner the random forest and the roles of covariates in the random forest can be explored interactively as the user highlights observations and groups of observations of interest.
\newline
\newline
The dashboards of \citet{Breiman2004} and \citet{Quach2012} both include parallel coordinates plots linked to the MDS plots.
These parallel coordinates plots assign each covariate a separate parallel vertical axis.
Separate parallel coordinates plots alternatively map to these axes the values of the covariates or the importance of these covariates to the prediction of individual observations.
Each observation is represented by a line that connects these parallel vertical axes.
In one of these types of parallel coordinates plots the vertical coordinate each line passes through at each axis indicates the value of the covariate represented by that axis associated with the observation represented by the line.
In the other type of parallel coordinates plot the vertical coordinate each line passes through at each axis indicates the importance of the associated covariate to the prediction of the observation represented by the line.
The points in the MDS plots and lines in the parallel coordinates plots may be coloured by one of a variety of attributes of the observations they represent.
These attributes include: the response class of the observation, whether the observation falls above or below a threshold in one covariate and whether or not the observation was correctly classified.
Points or lines selected on one plot are highlighted in all plots in the panel.
The interactive nature of these visualisations, whereby users can select subsets of observations from the MDS plots and have parallel coordinate plots of just these observation produced, allows the user to examine covariate effects within different subsets of the data.
Such subsets could be all the observations predicted to have a particular response class or all the observations correctly predicted to have a particular response class.
These parallel coordinates plots may be inspected for common patterns among lines representing observations of particular subsets of the data and contrasts between observations from different subsets of the data.
\newline
\newline
The RAFT includes a third use of parallel coordinates plots that facilitate insight into the roles of covariates in predicting individual response classes in random forests fitted for classification.
\citet{Breiman2004} refer to these plots as `prototypes'.
For a particular predicted response class, the observation is identified that has the most observations predicted to have this this same class among its nearest neighbours as defined by the proximity matrix for the random forest.
The distributions of the values of the covariates among these neighbouring observations are then summarised by intervals drawn on a parallel coordinates plot with axes for each covariate.
On the parallel vertical axis for each covariate the 25th percentile of the values of that covariate from the neighbourhood of observations is depicted as the lower bound of the interval, the median of these observations is depicted as the midline and the 75th percentile as the upper bound.
Three groups of line segments are drawn between the parallel vertical axes to form the intervals.
The first group of line segments connect the 25th percentiles on each axis, a second group connect the medians on each axis and a third group connect the 75th percentiles on each axis.
In this way the distributions of covariate values most closely associated with the prediction of the focal response class are represented for all covariates on a single parallel coordinate plot.
This plot is termed the `prototype' plot for that response class.
This procedure is repeated for successive response classes to create separate prototype plots.
Only the observations that have not yet been used to defined a prototype are considered to define the current prototype.
\newline
\newline
The distributions of covariate values associated with groups of observations identified from the proximity matrix of a random forest have also been represented with coxcomb plots.
\citet{Plonski2014} apply self organising maps (SOM) to the proximity matrix of a random forest to identify these groups.
They then depict the characteristic covariate values associated with each of these groups with a coxcomb plot in the array of coxcomb plots they produce.
This contrasts with the method of grouping observations into neighbourhoods to produce the prototypes plots of \citet{Breiman2004} though similar parallel coordinates plots of the distributions of covariate values could be produced for each of the groups of observations identified by the SOM based technique.
\newline
\newline
The visualisations reviewed above collectively facilitate deep insights into the roles of covariates in a random forest.
However, none of these visualisations communicate the identities and orders of covariates in the many hierarchies of interacting covariate effects that together constitute a random forest.
These hierarchies of covariate effects are defined by the paths from root to leaf nodes through the decisions trees of the random forest.
To further aid exploration of the roles of covariates in a random forest we propose novel applications of parallel coordinates plots and Sankey diagrams to represent all possible paths through the decision trees of a random forest.
We have designed these visualisations to communicate the identities and orders of the covariates that define these paths and to foreground the paths most frequently selected across the decision trees of a random forest.
In this way we present all of the sequences of interacting covariate effects that constitute a random forest in a single plot.
This plot also foregrounds the sequences of covariate effects that occur most frequently throughout the random forest.

\section{An example random forest}
\label{sec:example.rf}
We demonstrate our visualisations using a random forest fitted to some publicly available data for a classification problem.
The classification problem is that of ground cover classification from remotely sensed imagery.
These data consist of 6435 ground truthed pixels of satellite imagery.
The response variable is a categorical variable which discretises the type of ground cover present within these pixels into six classes.
Each observation of the ground cover class of a pixel is accompanied by the reflectance values of all pixels in a three pixel by three pixel rectangular grid centred on the pixel for which the ground cover class was observed.
The reflectance data for each pixel are available in four regions of the electromagnetic spectrum.
Thus there are 36 covariates.
These data are available from the University of California, Irvine Machine Learning Repository \citep{Newman1998} as the Statlog Landsat Satellite Data Set and are included in the R package `mlbench' \citep{mlbench2010}.
\newline
\newline
For our example we have used the the `randomForest' package \citep{randomForest2002} which is the reference implementation of random forests in the R language and environment for statistical computing \citep{R2017}.
We used the `caret' package \citep{caret} to construct 100 cross validation folds each with proportions of observations of the ground cover classes that that were similar to those present in the full data set.
We then performed 100 fold cross validation with the `caret' package to choose the size of the random sample of covariates that were available to define each binary partition during the model fitting process (the tuning parameter `mtry' in the `randomForest' package).
This procedure selected the value of eight for `mtry' to which we set this tuning parameter when we fitted a 500 tree random forest to the full data set.
The result was a random forest that correctly classified 75.7\% of the observations and had the covariate importance scores plotted in Supplementary Figure 1.

\section{Parallel Coordinates Plots of Paths}
\label{sec:pcp}
\begin{figure}
\begin{center}
\includegraphics[width=\textwidth]{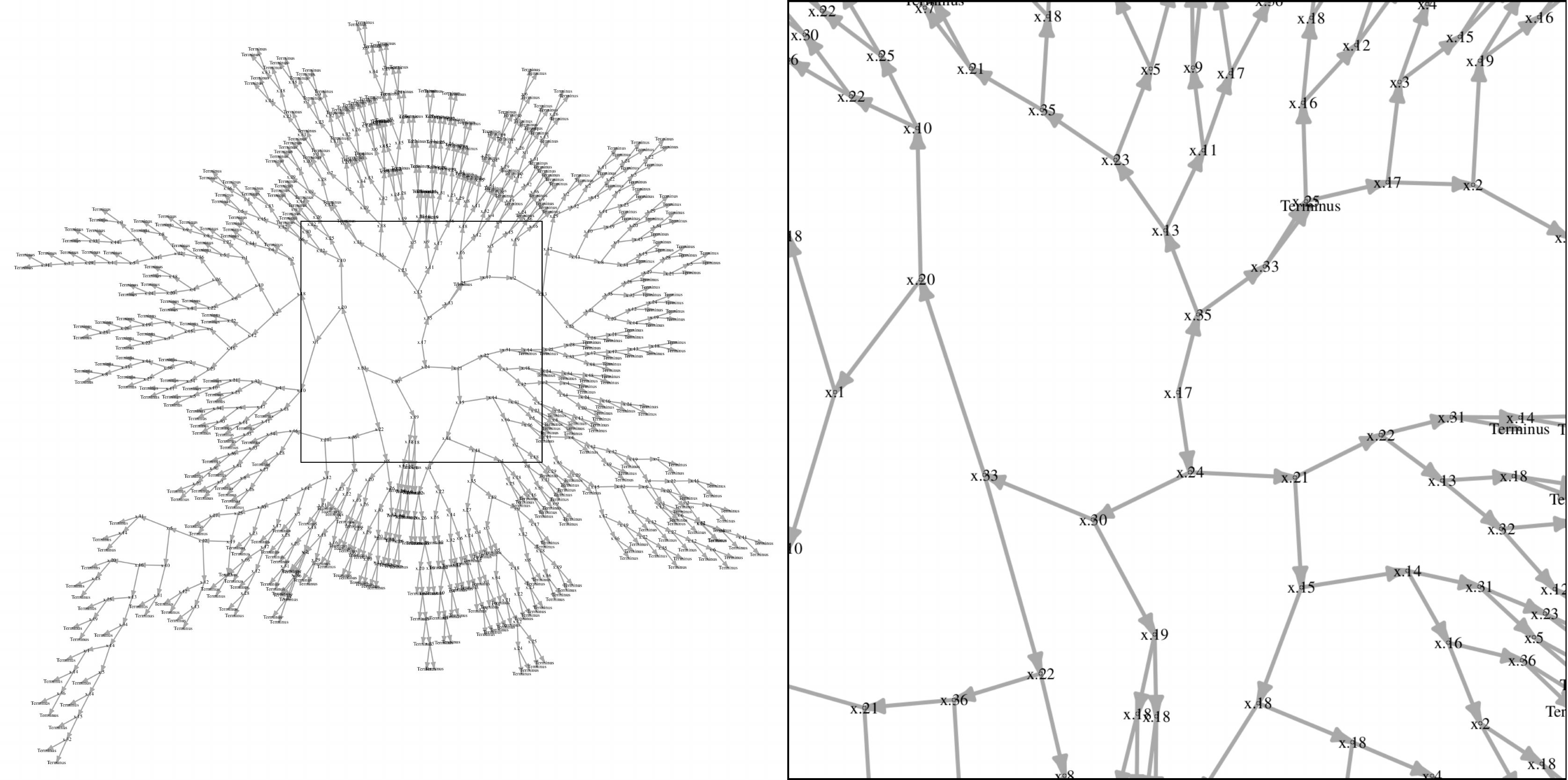}
\end{center}
\caption{A dendrogram representing a single decision tree extracted from our example random forest. Each node is labeled with the name of the covariate that defined that node. \label{fig:dend.1t}}
\end{figure}

\begin{figure}
\begin{center}
\includegraphics[width=\textwidth]{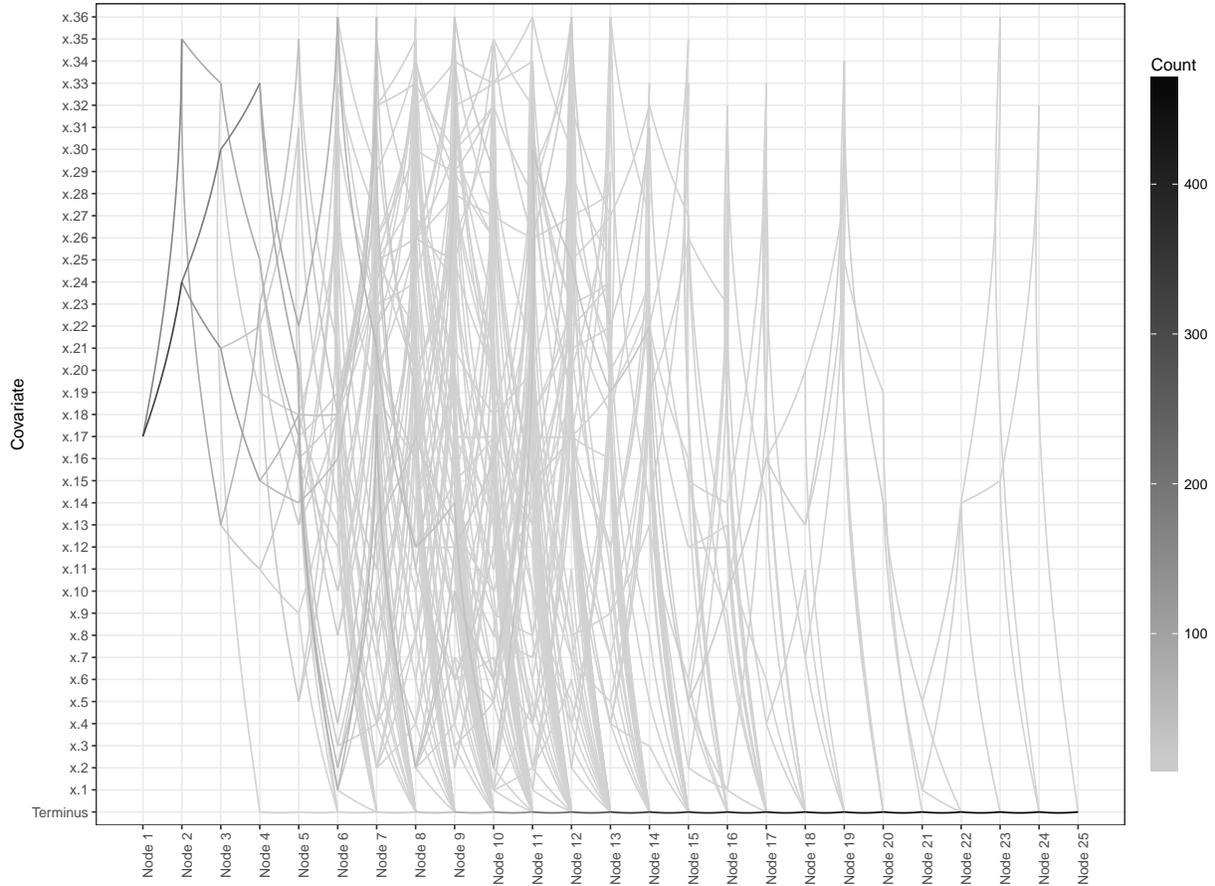}
\end{center}
\caption{A parallel coordinates plot depicting all paths through the decision tree depicted in Figure \ref{fig:dend.1t}. The sequences of nodes along the paths are represented by line segments connecting the parallel vertical axes. The covariates defining these nodes are represented by the gradations on these vertical axes. The number of edges between a pair of nodes is represented by the darkness of the line segment representing these edges. Node 1 is the root node and successive nodes along paths from this node to terminal nodes are represented by the successive parallel vertical axes to the right of the axis for Node 1. \label{fig:pc.1t}}
\end{figure}

\begin{figure}
\begin{center}
\includegraphics[width=\textwidth]{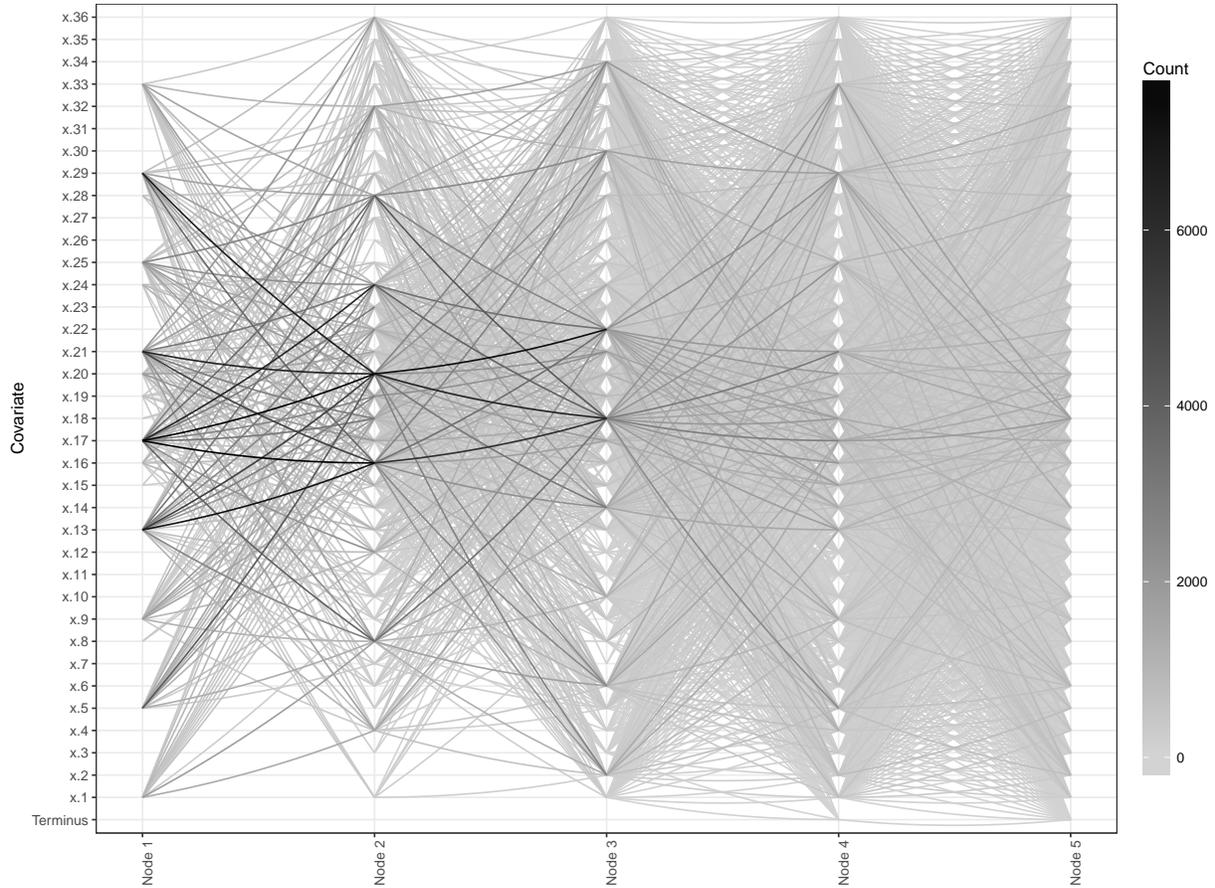}
\end{center}
\caption{A parallel coordinates plot depicting the first five nodes along all of the paths through all of the decision trees constituent to our example random forest. This parallel coordinates plot is arranged identically to that in Figure \ref{fig:pc.1t}. \label{fig:pc.at}}
\end{figure}

The hierarchies of interacting covariate effects that collectively constitute a random forest define the paths from root nodes to leaf nodes in the decision trees of that random forest.
We introduce our parallel coordinates plots of the paths through a random forest by first explaining how such plots can be created for all the paths through a single decision tree.
We then explain how this technique may be applied to visualise all the paths through a random forest composed of many such decision trees.
Figure \ref{fig:dend.1t} is a circular dendrogram representing the first tree of the 500 tree random forest we fitted to the ground cover classification data.
The left plot in Figure \ref{fig:dend.1t} depicts the full dendrogram with a square box outlining the section that has been magnified and reproduced in the right plot in this Figure.
The nodes are represented by labels which specify the covariate that defined that node.  
Edges are represented by arrows exiting each node and terminal nodes have each been labeled `Terminus'.
To produce this plot we extracted the first tree from the random forest and converted it into a directed network.
We then plotted this network with the `igraph' package  \citep{igraph}.
In this Figure the root node is located at the centre of the dendrogram and is labeled with the covariate which defined this node: `x.17'.
The leaf nodes, each labeled `Terminus', may be seen at the ends of the sequences of arrows emanating from the root node.
\newline
\newline
We also represent all the paths through the first tree of the random forest fitted to the ground cover classification data as a parallel coordinates plot in Figure \ref{fig:pc.1t}.
Once the tree was stored as a directed network the `igraph' package was used to determine the paths from the root node to each of the terminal `leaf' nodes.
Each of these paths is visible in the dendrogram in Figure \ref{fig:dend.1t}.
We then aggregated and transformed the data describing these paths into a format we could plot as a parallel coordinates plot using the `pairs' geometry from the R package `ggplot2' \citep{Wickham2009}.
This parallel coordinates plot is presented in Figure \ref{fig:pc.1t}.
The data aggregations and transformations necessary to produce each of the visualisations we introduce in this paper are performed with the aid of the R packages `dplyr' \citep{dplyr2016}, `tidyr' \citep{tidyr2017}, `purrr' \citep{purrr2016}, `magrittr' \citep{magrittr2016} and `forcats' \citep{forcats2017}.
\newline
\newline
In Figure \ref{fig:pc.1t} the paths from the root node to each of the terminal nodes are represented by sequences of line segments.
The orders of nodes along these paths are represented by the horizontal order of the parallel vertical axes from left to right.
The root node is represented by the left most vertical axis which is labeled `Node 1'.
Here we will refer to the number of nodes along a path from the root node to a particular node as the rank of that particular node.
The covariates defining nodes are represented by the gradations on these parallel vertical axes.
Terminal nodes of each rank are represented by the first gradation on each parallel vertical axis.
The tree represented by the parallel coordinates plot in Figure \ref{fig:pc.1t} has a root node defined by the covariate `x.17'.
Subsequently, the representations of each path through this tree originates on the vertical axis labeled Node 1 at the gradation representing the covariate `x.17'.
Two edges exit the root node, one of these edges connects to the node defined by the covariate `x.24' and the other connects to the node defined by covariate `x.35'.
These two edges are represented by line segments on the parallel coordinates plot between the representation of the root node at the position of the covariate `x.17' on the axis labeled Node 1 and the nodes represented by the positions of the covariates `x.24' and `x.35' on the axis labeled Node 2.
The darkness of a line segment in Figure \ref{fig:pc.1t} represents the number of paths through the decision tree that passed through a pair of nodes with ranks represented by the pair vertical axes connected by that line segment and defined by the covariates represented by the vertical coordinates on these axes which the line segment connects.
For example, the darkness of the left most pair of line segments in Figure \ref{fig:pc.1t} depicts how a larger proportion of the paths through the decision tree represented in this Figure commenced at the root node and passed through a second node defined by the covariate `x.24' than started at the root node and passed though a second node defined by the covariate `x.35'.
This feature of the decision tree may be corroborated by inspection of Figure \ref{fig:dend.1t}.
Figure \ref{fig:pc.at} is a parallel coordinates plot arranged identically to the parallel coordinates plot in Figure \ref{fig:pc.1t}.
In Figure \ref{fig:pc.at} we have represented the first five nodes along all of the paths through all of the 500 decision trees that constitute our example random forest.
From this plot it is apparent that the covariates most frequently selected to define the root nodes of these 500 decision trees are `x.17', `x.13', `x.21' and `x.29'.
Our software implementation of these parallel coordinates plots gives the user the option of producing a coloured version of these plots with a colour scale from the R package `viridis' \citep{viridis2016} replacing the grey scale used in Figure \ref{fig:pc.at}.

\section{Sankey Diagrams of Paths}
\label{sec:sankey}
\begin{figure}
\begin{center}
\includegraphics[height=0.69\textheight,angle = 90]{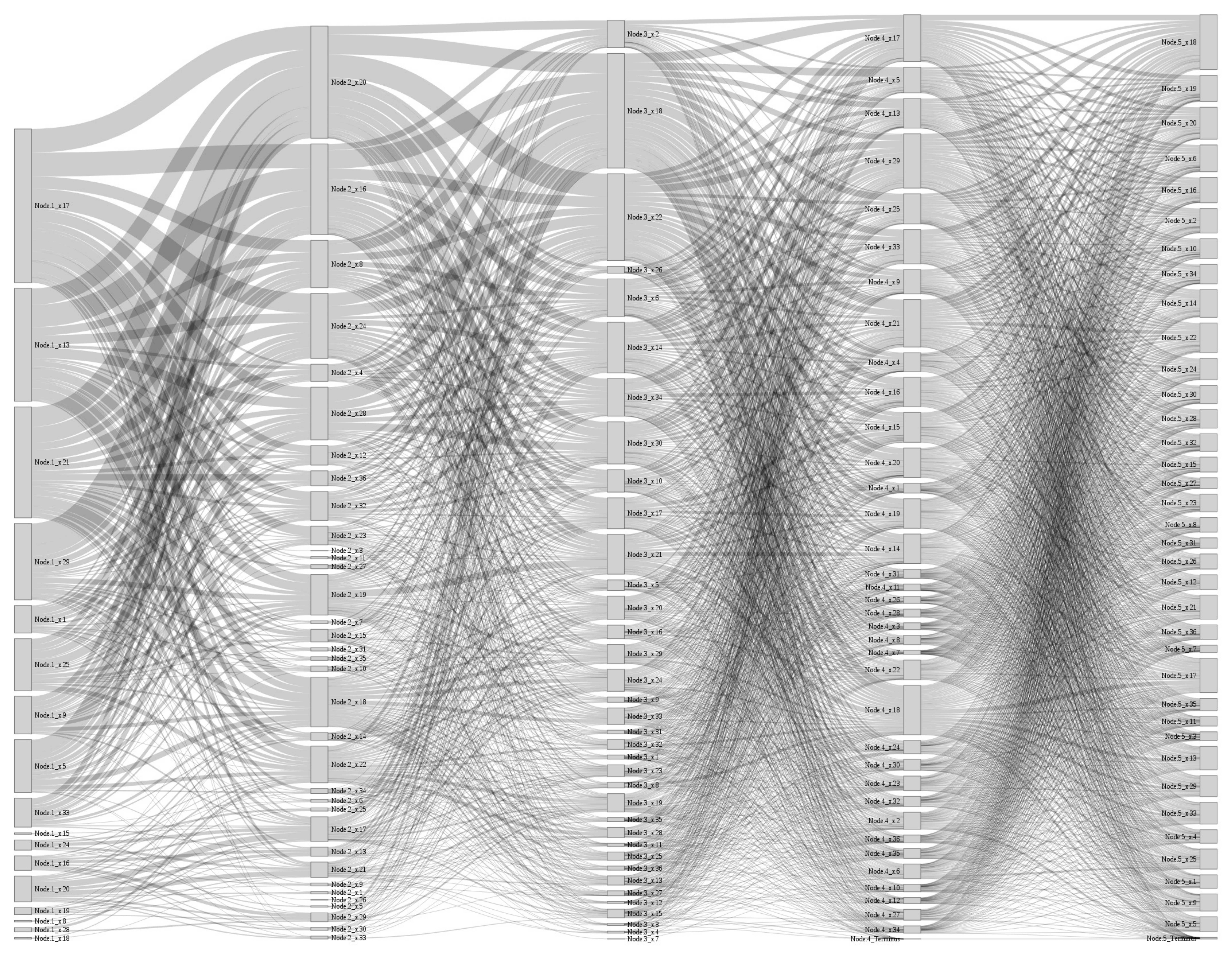}
\end{center}
\caption{A Sankey diagram depicting the first five nodes along all of the paths through all of the decision trees in our example random forest. The organisation of this diagram is similar to that of the parallel coordinates plots presented in Figure \ref{fig:pc.1t} and Figure \ref{fig:pc.at}.\label{fig:sk.at}}
\end{figure}

Sankey diagrams have been used to represent the flow of some quantity of interest between nodes in a network.
The nodes are represented as rectangular blocks and the flows between nodes are represented as curved links between the blocks.
The magnitude of a flow between nodes is represented by the width of the link representing this flow.
Conversion of the data used to generate Figure \ref{fig:pc.at} into a single directed network facilitated the representation of this information as a Sankey diagram with the R package `networkD3' \citep{networkD32017}. 
Figure \ref{fig:sk.at} displays a Sankey diagram of the first five nodes along all paths through all trees in our example random forest.
The organisation of this Sankey diagram is similar to that of the parallel coordinates plots in Figures \ref{fig:pc.1t} and \ref{fig:pc.at}.
In Figure \ref{fig:sk.at} the rectangular blocks represent groups of nodes in the decision trees of the random forest that were the same number of nodes along a path from a root node to a leaf node and were defined by the same covariate.
The blocks are organised into columns.
The sequence of columns from left to right represents the order of nodes along paths through the decision trees from root nodes to leaf nodes.
The left most column in this Sankey diagram contains blocks that represent the various root nodes of the decision trees.
The next column to the right contains blocks that represent the second nodes along the paths from root nodes to terminal nodes and so forth.
If we count the number of nodes along a path from the root node to a particular node this can be thought of as the rank of that node along the path.
In our Sankey diagrams each block is labeled with the rank of the nodes it represents and covariate that defines these nodes.
Thus a block labeled `Node.1\_x.17' represents all the root nodes defined by the covariate `x.17'.
The height of each block reflects the proportion of all the paths through all the trees that had a node at the position along a path represented by the horizontal position of the block where this node was defined by the covariate with which the block is labeled.
The width of a link between a pair of blocks represents the proportion of the paths through the trees in which an edge connected nodes with the characteristics encoded in the horizontal positions and labels of these blocks.
\newline
\newline
The `networkD3' package produces interactive visualisations that are displayed in a web browser and may also be written out as a HTML file. 
Examples of interactive Sankey diagrams produced with the R package that accompanies this paper have been included on the project website.
A link to this website is provided on the GitHub page that provides the R package (see Supplementary Materials).
These examples include a Sankey diagram of the paths through the trees of a random forest fitted to Anderson's Iris data \citep{Anderson1935}. 
These data are available in the base installation of R.
The Iris data contain a sufficiently small number of covariates that it was feasible to colour blocks and links by the identities of the associated covariates and still be able to visually distinguish these colours.
A static version of this Sankey diagram is included as Supplementary Figure 2.
Code to produce each of these Sankey diagrams is included among the examples in the R package.

\section{Discussion}
\label{sec:disc}
We have proposed visualisations that facilitate insights into the roles of covariates in a random forest additional to those available from the collection of visualisations that currently exists for this purpose.
Our novel applications of Sankey diagrams and parallel coordinates plots take a network flow approach to representing all possible paths through a random forest in manners that foreground the paths most frequently selected.
In this way, our visualisations emphasize the sequences of covariate interactions that are most important to the predictive mechanism of a random forest.
This visual communication of the identities and orders of covariates in these sequences along with the frequencies of these sequences has not been achieved in other visualisations of random forests to date.
Our visualisations can help a practitioner identify interesting pairs of covariates to examine with pairwise plots to assess interactions between covariates.
This will be particularly useful in situations where the number of covariates is large enough that inspection of a separate pairwise plot for every possible pairwise interaction would involve a substantial investment of time.
In this manner our plots complement the simple heatmaps of a single scalar metric of interaction strength for each potential pair of covariates allowing for more detailed inspection of the potential for important interactions across the whole forest.
Our plots also make apparent important interactions of substantially higher order than pairwise interactions which cannot be discovered from heatmaps of metrics for pairwise interaction strength.
\newline
\newline
Our visualisations complement parallel coordinates plots of covariate `prototypes' for each response class in classification problems.
Our plots depict the paths through a random forest in terms of the ordered hierarchies of covariate effects that define these paths.
In contrast, prototype plots represent the distributions of covariates from neighbourhoods of observations defined by the proximity matrix of the random forest.
Our visualisations display the orders of covariates in the hierarchies of interacting covariate effects involved in the predictive mechanism of a random forest while prototype plots do not display these orders.
In contrast, prototype plots display summaries of the values of covariates associated with prediction of particular response classes while our plots do not display these values.
\newline
\newline
The Sankey diagrams and parallel coordinates plots we propose in this paper also provide visual representations of the structure of the random forest as a whole.
The random forest fitted in our case study may be seen to have a few covariates that frequently define root nodes, several covariates that occasionally define root nodes and other covariates that rarely or never define root nodes.
This asymmetry in the proportions of nodes defined by each covariate at the roots of the trees may be seen to decrease steadily among groups of nodes increasing further from the root nodes.
By the fifth node along paths from the root nodes most of the covariates are defining nodes in approximately the same proportion of paths.
These visualisations also depict the proportion of nodes of each distance from the root node that are terminal nodes.
This provides another overall visual summary of the structure of the random forest as a whole.
Our visualisations of the structures of entire random forests would facilitate comparisons of groups of random forests.
Comparisons of groups of models are one of the recommendations \citet{Wickham2015}.
Groups of random forests it would be pertinent to compare would be those obtained from fitting random forests to the same data set using different values for the tuning parameters.  
\newline
\newline
In our case study comparisons of our visualisations to the scalar rankings of covariate importance plotted in Supplementary Figure 1 provides informative contrasts.
For example, the covariate that makes the greatest contribution to the predictive performance of the random forest is `x.18'.
The covariate `x.18' very rarely defines root nodes but is the most common covariate to define the third node along paths from root to leaf nodes.
Thus the contribution of `x.18' to the predictive performance of the random forest can be seen to be in no small part due to the interaction of this covariate with numerous other pairs of covariates that define the second and root nodes of the trees that constitute the random forest.
\newline
\newline
Of the two styles of visualisations we have proposed in this paper, we find the Sankey diagrams more informative.
The Sankey diagrams convey the relative frequencies with which nodes of a particular distance from the root node were defined by different covariates through the relative heights of the rectangular blocks representing these groups of nodes.
This information is not readily available from the parallel coordinates plots and may only be very approximately inferred from the number and darkness of line segments exiting the representation of groups of nodes of a particular rank that were defined by a particular covariate.
In the interactive Sankey diagrams a user may hover the mouse over a rectangular block that represents a group of nodes and have all the links that represent edges connecting to those nodes highlighted.
This ability to foreground a particular feature of the Sankey diagram is very useful given the volume of information that is being represented in these plots.
Furthermore, highlighting a link results in a text box being displayed.
The text box identifies the covariates which define the nodes connected by the edges represented by this link.
The text box also specifies the frequency of edges between such nodes across the random forest.
Both the readily available interactivity and the additional information available visually from the Sankey diagrams  lead us to favour these over the parallel coordinates plots as a method for visualising the roles of covariates in a random forest.
While our visualisations have been developed for random forests they could be applied to other techniques that utilise one or more decision trees.
All paths through a CART \citep{Breiman1984} could be depicted with our techniques.
Our techniques could also be extended to visualise the paths through the decision trees of other methods based on ensembles of decision trees such as Gradient Boosted Machines \citep{Friedman2001}.
\newline
\newline
One relatively straightforward elaboration upon our visualisations of random forests for classification would be to produce separate visualisations for each of the predicted response classes. 
This would involve the production of a panel of visualisations.
Each visualisation in the panel would represent all paths through the random forest that lead to prediction of a particular response class.
Such panels could accompany an overall visualisation of all the paths and would be of particular interest in situations where there was substantial inequality among the numbers of observations of different response classes.
These response class specific visualisations could also suggest revealing horizontal orders for the vertical axes in the associated prototype plots.
Another extension of our visualisations could involve giving users the option to apply some threshold so that only the most common nodes at each position along the paths were displayed in the plot.
A user could then experiment with values for this threshold to produce visual summaries of the random forest of varying granularity.
This would be particularly useful for visualising random forests constructed from large sets of covariates.
Further interactivity could be added whereby clicking a particular block or link produced additional visualisations of the characteristics of the random forest represented by that block or link.
For example clicking a particular block could produce a visualisation of the empirical probability density of the threshold values of the covariate that defines all the nodes represented by this block.
Another example could involve clicking a link between two nodes resulting in the production of visualisations pertinent to investigation of the nature of the interaction between the covariates defining these two nodes.
For instance, an ICE plot could be produced for one of these covariates coloured by the second covariate.
The observations that were passed down this link could be drawn as points on this ICE plot.
Development of such interactive `dashboard' style visualisations could be attempted with a combination of technologies such as R, `plotly' and `Shiny' but would be a substantial project.

\bigskip
\begin{center}
{\large\bf SUPPLEMENTARY MATERIAL}
\end{center}

\begin{description}

\item[R-package:] An R package implementing our methods for visualising the roles of covariates in random forests is provided at \url{https://github.com/brfitzpatrick/forestviews}.

\item[Supplementary Figure 1] A dot plot of the covariate importance scores obtained from our example random forest fitted to the ground cover classification data.
  
\item[Supplementary Figure 2] A Sankey diagram of the first five nodes of all paths through a random forest fitted to Anderson's Iris data. Colour depicts the identities of the covariates associated with the different elements of this diagram.

\end{description}

\end{document}